\documentclass[a4paper,fleqn,usenatbib,useAMS]{mnras}

\usepackage{graphicx,times}
\usepackage{amsmath}	
\usepackage{amssymb}	
\usepackage{multicol}        
\usepackage{bm}		
\usepackage{pdflscape}	
\usepackage{color}
\usepackage{float}
\usepackage{multirow}
\usepackage[T1]{fontenc}
\usepackage{ae,aecompl}

\title[A time-dependent particle acceleration and emission model]{A time-dependent particle acceleration and emission model: Understanding the particle spectral evolution and blazar flares}

\author[Zheng et al.]{
Y. G. Zheng,$^{1,2,3}$\thanks{E-mail: ynzyg@ynu.edu.cn}
S. J. Kang,$^{4}$
C. Y. Yang,$^{2,3}$
J. M. Bai,$^{2,3}$\thanks{E-mail: baijinming@ynao.ac.cn}
\\
$^{1}$Department of Physics, Yunnan Normal University, Kunming, 650092, China\\
$^{2}$Yunnan Observatories, Chinese Academy of Sciences, Kunming 650011, China\\
$^{3}$Key Laboratory for the Structure and Evolution of Celestial Objects, Chinese Academy of Sciences\\
$^{4}$School of Physics and Electrical Engineering, Liupanshui Normal University, Liupanshui, Guizhou, 553004, China\\
}

\date{Accepted 2020 September 17. Received 2020 September 15; in original form 2020 April 9}

\pubyear{2020}

\begin{document}
\label{firstpage}
\pagerange{\pageref{firstpage}--\pageref{lastpage}}
\maketitle

\begin{abstract}

The jets of blazars are renowned for their multi-wavelength flares and rapid extreme variability; however, there are still some important unanswered questions about the physical processes responsible for these spectral and temporal changes in emission properties. In this paper, we develop a time-dependent particle evolution model for the time-varying emission spectrum of blazars. In the model, we introduce time-dependent electric and magnetic fields, which consistently include the variability of relevant physical quantities in the transport equation. The evolution on the electron distribution is numerically solved from a generalized transport equation that contains the terms describing the electrostatic, first-order and second-order \emph{Fermi} acceleration, escape of particles due to both advection and spatial diffusion, as well as energy losses due to the synchrotron emission and inverse-Compton scattering of both synchrotron and external ambient photon fields. We find that the light curve profiles of blazars are consistent with the particle spectral evolution resulting from time-dependent electric and magnetic fields, rather than the effects of the acceleration or the cooling processes. The proposed model is able to simultaneously account for the variability of both the energy spectrum and the light curve profile of the BL Lac object Mrk 421 with reasonable assumptions about the physical parameters. The results strongly indicate that the magnetic field evolution in the dissipated region of a blazar jet can account for the variabilities.

\end{abstract}




\section{Introduction} \label{sec:intro}

Blazars are the radio-loud active galactic nuclei (AGN) characterized by non-thermal continuum emissions at all observed wavelengths, rapid variability, and a high degree of polarization (e.g., \citealt{1981ApJ...243...60M}). It is well-known that in the context of the unified AGN model (\citealt{1995PASP..107..803U}), these observational properties can be consistently interpreted with their relativistic plasma jets pointed nearly towards the observer, resulting in the substantial Doppler boosting of the emission spectra (\citealt{1979ApJ...232...34B}). The Doppler enhancement means that the emission spectra from blazars is dominated by jet dissipated region making the spectra invaluable for understanding jet physics and high energy particle acceleration. We can directly utilize the spectral and temporal changes in emission properties to understand the particle acceleration and dissipation mechanisms operating in jets.

There are some blazars that appear with noticeable and rapid flux variability, while the very fast variability events are resolved in as short as a few minutes and dramatically increase in luminosity (e.g., \citealt{2007ApJ...664L..71A,2007ApJ...669..862A}). Motivated by the nature of their variability, the unprecedented study of multi-wavelength observation campaign has emerged (e.g., \citealt{2011ApJ...729....2A,2011ApJ...738...25A,2011MNRAS.410..368B,2015A&A...578A..22A,2016ApJ...819..156B,2016ApJS..222....6B,2016ApJ...827...55K,2017ApJ...834....2A,2018A&A...620A.181A}),
where the spectra of blazars are obtained as close in time as possible. Consequently, strong evidence for the variable behaviors of energy spectra during the flaring epochs (e.g., \citealt{2011ApJ...736..131A,2018A&A...620A.181A}), distinct from the early observations, has been detected. On the other hand, the variability of some flaring events exhibit largely stochastic behavior, without clear correlation between variabilities at different frequencies (e.g., \citealt{2009A&A...502..749A,2012A&A...542A.100A}), whilst the variability of some other flaring events show a distinct light curve profile at observed wavelengths, sometimes with energy dependent time lags
(e.g., \citealt{2000ApJ...541..153F,2001ApJ...563..569T,2002ApJ...572..762Z,2004A&A...424..841R,2007ApJ...669..862A,2008ApJ...680L...9S,2010ApJ...710..810A}).

Previous modeling efforts to study temporal spectral behavior from a blazar have been concentrated mainly on the production of variability via either time-dependent particle distribution (e.g., \citealt{1996ApJ...461..657B,1999MNRAS.306..551C,2002ApJ...581..127B,2006A&A...453...47K}; \citealt{2013ApJ...764..113Z,2014MNRAS.442.3166Z,2015A&A...573A...7W}) or physical characteristics (e.g., \citealt{1979ApJ...232...34B,1985ApJ...298..114M,1991ApJ...377..403C,1997A&A...320...19M,1998A&A...333..452K,2007A&A...462...29G,2008A&A...491L..37M}) and/or geometrical parameters changing (e.g., \citealt{1992A&A...255...59C,1992A&A...259..109G}). These paradigms are able to successfully reproduce snapshots or time averaged spectral energy distributions. Although the multi-zone models are developed to understand the erratic nature of the variability and polarisation in blazars (\citealt{2014ApJ...780...87M,2016ApJ...829...69Z}), the issue of utilizing the information that is encoded in the observed time evolutions of the spectral energy distributions remains open. The model proposed in this work introduce time-dependent electric and magnetic (EM) fields into the transport equation, which allows the incorporation of other physical quantities' variabilities with respect to time. Our conjecture is that the magnetic field evolution in the dissipation region of a blazar jet can account for the variability.

The present paper is organized as follows. In Section 2, we deduce the transport equation that contains the terms of the first-order \emph{Fermi} acceleration, second-order \emph{Fermi} acceleration, particle escape, and energy losses. The section includes a description of the time-dependent parameters. In Section 3, we describe the ambient photon field, which contains the synchrotron photons of the same population of electrons and the photon sources external to the jet. In Section 4, we introduce an exponential time-profile-function to substantiate the time-dependent model parameters. In Section 5, we evolve the particle distributions in the case of constant physical parameters and time-dependent physical parameters to determine if the timescales of the particle spectra evolution achieve a steady state. In Section 6, we test the model's expected variability of the energy spectrum and the light curve profile. Then, we apply the model to explain the activity of Mrk 421, and some discussions are given in Section 7. Throughout the paper, we assume the Hubble constant $H_{0}=75$ km s$^{-1}$ Mpc$^{-1}$, the dimensionless numbers for the energy density of matter $\Omega_{\rm M}=0.27$, the dimensionless numbers of radiation energy density $\Omega_{\rm r}=0$, and the dimensionless cosmological constant $\Omega_{\Lambda}=0.73$. All of the physical quantities are calculated in a coordinate system co-moving with the radiation source.
\section{Basic equations}
Due to the random motion of magnetized plasma elements in a turbulent and tenuous plasma carrying a magnetic field, the momentum distribution of energetic particles that experience the \emph{Fermi} type acceleration can be described in terms of a diffusion equation in momentum space (\citealt{1967JETP...25..317T,1977MAtom.........T})

\begin{equation}
\frac{\partial f(p,t)}{\partial t}=\frac{1}{p^{2}}\frac{\partial}{\partial p}\biggl[p^{2}D(p,t)\frac{\partial f(p,t)}{\partial p}\biggr]+Q(p,t),
\label{Eq:1}
\end{equation}
where $f(p,t)$ is the isotropic, homogeneous phase space density in the unit of $\rm p^{-3}~cm^{-3}$ at the time $t$, $p$ the particle momentum, $D(p, t)=D(t)m_{e}cp$ the second order \emph{Fermi} acceleration diffusion coefficient of scattering MHD waves with a time-dependent momentum diffusion rate $D(t)$, $m_{e}$ is the electron rest mass, $c$ is the speed of light, and $Q(p,t)$ are the sources and sinks of particles in the unit of $\rm p^{-3}~cm^{-3}~s^{-1}$. Incorporating the effects of gaining momentum by a first order \emph{Fermi} acceleration from the strong shocks (\citealt{1978MNRAS.182..147B}); turbulence and/or electrostatic acceleration around the shock; both the synchrotron and the inverse Compton (IC) scatter cooling processes; and the escape of particles into Eq. {\ref{Eq:1}}, we obtain

\begin{eqnarray}
\frac{\partial f(p,t)}{\partial t}&=&\frac{1}{p^{2}}\frac{\partial}{\partial p}\biggl\{p^{2}\biggl[D(p,t)\frac{\partial f(p,t)}{\partial p}-\dot{p}_{\rm gain}(t)f(p,t)\nonumber\\&-&\dot{p}_{\rm loss}(t)f(p,t)\biggr]\biggr\}-\frac{f(p,t)}{t_{\rm esc}(p,t)}+Q(p,t)\,,
\label{Eq:2}
\end{eqnarray}
where $\dot{p}_{\rm gain}(t)$ is the first-order momentum gain rate at the time $t$ in the unit of $\rm p~s^{-1}$, $\dot{p}_{\rm loss}(t)$ is the momentum loss rate at the time $t$ in the unit of $\rm p~s^{-1}$, and $t_{\rm esc}(p,t)$ is the momentum-dependent escape timescale at the time $t$. Assuming an isotropic distribution of particles in the momentum space, the phase space density is related to the particle number density, $N(p,t)$, via $N(p,t)=4\pi p^{2}f(p,t)$. Therefore, we can rebuild the basic transport equation as
\begin{eqnarray}
\frac{\partial N(p,t)}{\partial t}&=&\frac{\partial}{\partial p}\biggl\{D(p,t)\frac{\partial N(p,t)}{\partial p}+\biggl[-\frac{2D(p,t)}{p}-\dot{p}_{\rm gain}(t) \nonumber\\&-&\dot{p}_{\rm loss}(t)\biggr]N(p,t)\biggr\}-\frac{N(p,t)}{t_{\rm esc}(p,t)}+4\pi p^{2}Q(p,t)\,.
\label{Eq:3}
\end{eqnarray}
It is convenient to transform the variable $p$ to the Lorentz factor, $\gamma$, by the relationship $N(p,t)dp=N(\gamma,t)d\gamma$ with $p=\gamma m_{e}c$. The transport equation in the $(\gamma,t)$ space can be written as
\begin{eqnarray}
\frac{\partial N(\gamma,t)}{\partial t}&=&\frac{\partial}{\partial \gamma}\biggl\{D(t)\gamma\frac{\partial N(\gamma,t)}{\partial \gamma}+\biggl[-2D(t)-\dot{\gamma}_{\rm gain}(t) \nonumber\\&-&\dot{\gamma}_{\rm loss}(t)\biggr]N(\gamma,t)\biggr\}-\frac{N(\gamma,t)}{t_{\rm esc}(\gamma,t)}+Q(\gamma,t)\,,
\label{Eq:4}
\end{eqnarray}
where the injection of the particles is described by $Q(\gamma,t)=4\pi m_{e}cp^{2}Q(p,t)$.

The evolution of the particle distributions requires the parameterization of the time-dependent terms of diffusion, energy gain, energy loss, escape, and injection. Following the early efforts of \cite{2019ApJ...873....7Z}, we can rewrite the momentum diffusion rate as
\begin{equation}
D(t)=\frac{eB(t)\sigma_{\rm mag}(t)}{3\eta(t) m_{e}c}\,,
\label{Eq:5}
\end{equation}
where $e$ is the charge of an electron, $B(t)$ is the time-dependent local magnetic field strength, $\sigma_{\rm mag}(t)$ is the magnetization parameter, and $\eta(t)$ is the time-dependent dimensionless parameter to represent the mean-free path relative to the Larmor radius (e.g., \citealt{2016ApJ...833..157K,2019ApJ...873....7Z}). The parameter $\dot{\gamma}_{\rm gain}(t)$ describes the electrostatic and shock acceleration of the particles
\begin{equation}
\dot{\gamma}_{\rm gain}(t)=\dot{\gamma}_{\rm elec}(t)+\dot{\gamma}_{\rm sh}(t)\,,
\label{Eq:6}
\end{equation}
where, $\dot{\gamma}_{\rm elec}(t)$ is the electrostatic acceleration rate with
\begin{equation}
\dot{\gamma}_{\rm elec}(t)=\frac{eE(t)}{m_{e}c}\,,
\label{Eq:7}
\end{equation}
and $\dot{\gamma}_{\rm sh}(t)$ the shock acceleration rate with
\begin{equation}
\dot{\gamma}_{\rm sh}(t)=\frac{3eB(t)\xi(t)}{4m_{e}c}\,.
\label{Eq:8}
\end{equation}
Here, $E(t)$ is the time-dependent electric field of strength, and $\xi(t)$ is the efficiency factor at time $t$. The parameter $\dot{\gamma}_{\rm loss}(t)$ describes the synchrotron and IC scatter cooling of the particles
\begin{equation}
-\dot{\gamma}_{\rm loss}(t)=\dot{\gamma}_{\rm syn}(t)+\dot{\gamma}_{\rm ic}(t)\,.
\label{Eq:9}
\end{equation}
where, $\dot{\gamma}_{\rm syn}(t)$ is the rate of synchrotron cooling with
\begin{equation}
\dot{\gamma}_{\rm syn}(t)=\frac{4\sigma_{T}u_{B}(t)}{3m_{e}c}\gamma^{2}=\frac{\sigma_{T}B^{2}(t)}{6\pi m_{e}c}\gamma^{2}\,,
\label{Eq:10}
\end{equation}
and $\dot{\gamma}_{\rm ic}(t)$ is the rate of IC scatter cooling in the ambient photon fields, $u_{\rm ph}(\gamma,t)$, with
\begin{equation}
\dot{\gamma}_{\rm ic}(t)=\frac{4\sigma_{T}u_{\rm ph}(\gamma,t)}{3m_{e}c}\gamma^{2}=3.25\times10^{-8}u_{\rm ph}(\gamma,t)\gamma^{2}~~\rm s^{-1}\,.
\label{Eq:11}
\end{equation}
Here, $\sigma_{T}$ is the Thomson cross section, $u_{B}(t)=B^{2}(t)/8\pi$ the magnetic field energy density.

The escape timescale of the particles
\begin{equation}
t_{\rm esc}(\gamma,t)=\biggl[\frac{C(t)}{\gamma}+F(t)\gamma\biggr]^{-1}\,,
\label{Eq:12}
\end{equation}
where, $C(t)$ is the shock regulated escape rate at time $t$ with
\begin{equation}
C(t)=\frac{eB(t)}{\omega(t)m_{e}c}\,,
\label{Eq:13}
\end{equation}
and $F(t)$ the \emph{Bohm} diffusive escape rate at time $t$ with
\begin{equation}
F(t)=\frac{\eta(t) m_{e}c^{3}}{r_{s}^{2}e B(t)}\,.
\label{Eq:14}
\end{equation}
Here, $\omega(t)$ is a time-dependent dimensionless parameter with the order of unity that accounts for the time dilation and obliquity in the relativistic shock (e.g., \citealt{2016ApJ...833..157K}), and $r_{s}$ is the size of the blob. The particle injection mechanism is not well understood, but we assume here that the electrons accelerated inside the blob originating as members of a low-energy thermal distribution, and that the acceleration and emission regions are co-spatial. In the calculations presented in this work, we define the function of the injection rate $Q(\gamma, t)$ separately for each evolution test.

\section{Ambient Photon Field}
It is noted that if the high energy relativistic electrons and/or positrons interact with an ambient photon field, the photons can be inverse Compton (IC) scattered to very high energies. Such photon fields can be either the synchrotron photons of the same population of electrons (SSC) or the photon sources external to the jet (EC); in some special cases, both. Since we intend to calculate the evolution of the particle distributions, these types of energy transfers between the electrons and photons in the energy loss rate of relativistic electrons must be included. In this section, we establish the two photon fields. The context assumes that the relativistic plasma propagates with an associated bulk Lorentz factor $\Gamma$ in a stationary spherical blob.
\subsection{Synchrotron Photon Field}
For the spatially isotropically distributed relativistic electrons, the generated synchrotron photons are spatially isotropically distributed with the following differential number density
\begin{equation}
n(\epsilon_{\rm syn}, t)=\frac{4\pi}{hc\epsilon_{\rm syn}}\frac{j_{\rm syn}(\nu, t)}{k_{\rm syn}(\nu, t)}[1-e^{-k_{\rm syn}(\nu, t)r_{\rm s}}]~~\rm cm^{-3}\;,
\label{Eq:15}
\end{equation}
where, $h$ is the Planks constant, $\epsilon_{\rm syn}$ is the synchrotron photon energies in the unit of $m_{e}c^{2}$, $j_{\rm syn}(\nu, t)$ is the isotropic synchrotron emissivity, $k_{\rm syn}(\nu, t)$ is the absorption coefficient, and $r_{\rm s}$ is the radius of the spherical blob. Using the synchrotron photon density, we can calculate the synchrotron photon field as
\begin{equation}
u_{\rm syn}(\gamma, t)=m_{e}c^{2}\int\epsilon_{\rm syn}n(\epsilon_{\rm syn}, t)d\epsilon_{\rm syn}~~\rm erg~cm^{-3}\;.
\label{Eq:16}
\end{equation}

We note that the Klein-Nishina (KN) effect can modify the electron distribution of extreme high energy tails
(e.g., \citealt{2005MNRAS.363..954M,2009ApJ...703..675N}).
Taking this effect into account, we can rewrite Eq. (\ref{Eq:16}) as
\begin{equation}
u_{\rm syn, KN}(\gamma, t)=m_{e}c^{2}\int f_{\rm KN}(\kappa)\epsilon_{\rm syn}n(\epsilon_{\rm syn}, t)d\epsilon_{\rm syn}~~\rm erg~cm^{-3}\;,
\label{Eq:17}
\end{equation}
with $\kappa=4\gamma\epsilon_{\rm syn}$. Where, the function $f_{\rm KN}(\kappa)$ can be approximated as follows (\citealt{2005MNRAS.363..954M}):
\begin{equation}
f_{\rm KN}(\kappa)\simeq\left\{ \begin{array}{ll}
1          & ~\kappa\ll 1 ~\mbox{(Thomson limit)}\\
\frac{9}{2\kappa^2}(\ln\kappa-\frac{11}{6})     & ~ \kappa\gg 1~ \mbox{(KN limit)}\;.
\end{array} \right.
\label{Eq:18}
\end{equation}
When it satisfies $\kappa\leq 10^{4}$, $f_{\rm KN}(\kappa)\simeq 1/(1+\kappa)^{3/2}$.

\subsection{External Photon Field}
The energy density of an external photon field is associated with the distance between the position of the dissipation region and the central supermassive black hole (e.g., \citealt{2012ApJ...754..114H,2014ApJ...782...82D,2016MNRAS.457.3535Z,2017ApJS..228....1Z}). In the present work, we simply assume a constant isotropic photon field in the galactic frame (e.g., \citealt{2012ApJ...761..110Z}) as
\begin{equation}
u^{\prime}_{\rm ex}=\frac{L^{\prime}_{d}\tau_{d}}{4\pi R^{\prime2}_{d}c}\,,
\label{Eq:19}
\end{equation}
where, $L^{\prime}_{d}$ is the luminosity of the accretion disk surrounding the central supermassive black hole, $\tau_{d}$ is the scattering depth of the ambient medium scattering the accretion disk photons, and $R^{\prime}_{d}$ is the radius up to where the accretion disk photons are scattered by the ambient medium. Transforming the coordinate system from galactic to jet, we obtain an isotropically distributed external photon field (e.g., \citealt{1993ApJ...416..458D})
\begin{eqnarray}
u_{\rm ex}&=&u^{\prime}_{\rm ex}\Gamma^{2}(\frac{4}{3}-\frac{1}{3\Gamma^{2}})\nonumber\\&\simeq&\frac{\Gamma^{2}L^{\prime}_{d}\tau_{d}}{3\pi R^{\prime2}_{d}c}\,.
\label{Eq:20}
\end{eqnarray}
Combining Eqs. (\ref{Eq:17}) and (\ref{Eq:20}), we can obtain the ambient photon fields, $u_{\rm ph}(\gamma, t)$, appearing in Eq. (\ref{Eq:11})
\begin{equation}
u_{\rm ph}(\gamma, t)=u_{\rm syn, KN}(\gamma, t)+u_{\rm ex}\,.
\label{Eq:21}
\end{equation}
\section{Time Profile Function}
\begin{figure}
	\centering
		\includegraphics[height=6cm,width=8cm]{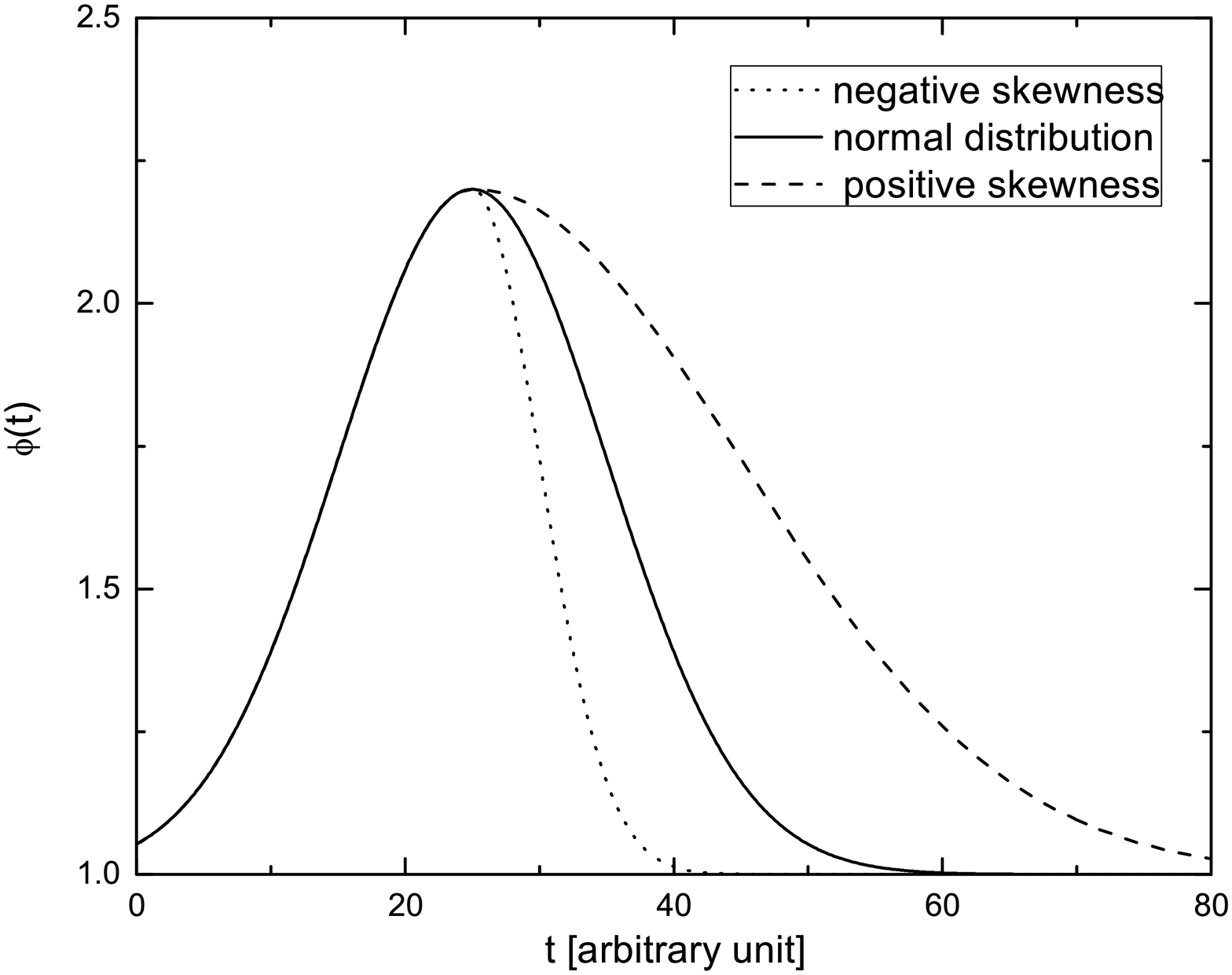}
		    	\caption{Corresponding profiles for a given time, $t_{*}=25$, with an arbitrary time unit. The plots include a comparison of the effects of changes in the parameter $\theta_{2}$. The parameters $\alpha=1.2$ and $\theta_{1}=10.0$ with arbitrary time unit remain constant. Three values of $\theta_{2}$ are adopted: $\theta_{2}=5.0$ (dotted curve), $\theta_{2}=10.0$ (solid curve), and $\theta_{2}=20.0$ (dashed curve).}
	\label{Fig:1}
\end{figure}

In order to simulate the variability of the energy spectra and the light curve, we assume that the physical variations of the parameters are correlated in time. Since the Gaussian profile is a reasonable approximation to the time series in the field of blazar physics (\citealt{2017Galax...5...19K}), we introduce an exponential time-profile-function, $\phi(t)$, to describe the time-dependent modulations of the EM fields (e.g., \citealt{2018ApJ...853...16K,2019ApJ...872...65K}) due to the impulsive reconnection occurring at the shock. Such that
\begin{equation}
\phi\left(t\right)=\left\{ \begin{array}{l}
1+\alpha e^{-\frac{(t-t_{*})^{2}}{2\theta_{1}^{2}}}\;\;\;\;\;\;\;t< t_{*}\,,~~\theta_{1}\ll t_{*}\\
1+\alpha e^{-\frac{(t-t_{*})^{2}}{2\theta_{2}^{2}}}\;\;\;\;t \ge t_{*}
\end{array} \right.\;,
\label{Eq:22}
\end{equation}
where $\alpha$ is a dimensionless constant, $\theta_{1}$ and $\theta_{2}$ are the width of the root mean square for the time-profile-function, and $t_{*}$ is the given time of the peak intensity for the time-profile-function. The corresponding profiles for a given time of peak intensity are plotted in Figure {\ref{Fig:1}}. The plots include a comparison of the effects of changes in the parameters $\alpha$ and $\theta_{2}$. Two important observations are made based on this figure. 1) the function $\phi(t)$ increases from the initial value $\phi(0)\simeq1$, and reaches a maximum value at the given time $t_{*}$; 2) either the symmetrical or asymmetrical profile can be expected by changing the parameter $\theta_{2}$. It shows a negative skewness distribution profile when the parameter satisfies $\theta_{2}<\theta_{1}$, a normal distribution profile when the parameter satisfies $\theta_{2}=\theta_{1}$, and a positive skewness distribution profile when the parameter satisfies $\theta_{2}>\theta_{1}$.

Following the approach of \cite{2018ApJ...853...16K}, we postulate the momentum diffusion rate $D(t)=D_{0}\phi(t)$, the rate of the both electrostatic and shock acceleration $\dot{\gamma}_{\rm gain}(t)=\dot{\gamma}_{\rm gain,0}\phi(t)$, the rate of synchrotron cooling $\dot{\gamma}_{\rm syn}(t)=\dot{\gamma}_{\rm syn,0}\phi(t)$, the shock regulated escape rate $C(t)=C_{0}\phi(t)$, the \emph{Bohm} diffusive escape rate $F(t)=F_{0}\phi(t)$, and the injection of the particles $Q(\gamma, t)=Q_{0}(\gamma)\phi(t)$. Here, the subscript ``0'' denotes the initial value of a quantity at the beginning of the evolution. We can relate Eqs. (\ref{Eq:5}), (\ref{Eq:6}), (\ref{Eq:7}), (\ref{Eq:8}), (\ref{Eq:10}), (\ref{Eq:13}), and (\ref{Eq:14}) to obtain the initial momentum diffusion rate as follows
\begin{equation}
D_{0}=\frac{eB_{0}\sigma_{\rm mag,0}}{3\eta_{0} m_{e}c}=5.86\times10^{5}\sigma_{\rm mag,0}\eta_{0}^{-1}\biggl(\frac{B_{0}}{0.1~\rm G}\biggr)~~\rm s^{-1}\,,
\label{Eq:23}
\end{equation}
the initial rate of the both electrostatic and shock acceleration
\begin{equation}
\dot{\gamma}_{\rm gain,0}=\dot{\gamma}_{\rm elec,0}+\dot{\gamma}_{\rm sh,0}\,,
\label{Eq:24}
\end{equation}
with the initial rate of electrostatic acceleration
\begin{equation}
\dot{\gamma}_{\rm elec,0}=\frac{eE_{0}}{m_{e}c}=1.76\times10^{6}\biggl(\frac{E_{0}}{B_{0}}\biggr)\biggl(\frac{B_{0}}{0.1~\rm G}\biggr)~~\rm s^{-1}\,,
\label{Eq:25}
\end{equation}
and the initial rate of shock acceleration
\begin{equation}
\dot{\gamma}_{\rm sh,0}=\frac{3eB_{0}\xi_{0}}{4m_{e}c}=1.32\times10^{6}\xi_{0}\biggl(\frac{B_{0}}{0.1~\rm G}\biggr)~~\rm s^{-1}\,,
\label{Eq:26}
\end{equation}
the initial rate of synchrotron cooling
\begin{equation}
\dot{\gamma}_{\rm syn,0}=\frac{\sigma_{T}B_{0}^{2}}{6\pi m_{e}c}\gamma^{2}=1.29\times10^{-9}B_{0}^{2}\gamma^{2}~~\rm s^{-1}\,,
\label{Eq:27}
\end{equation}
the initial shock regulated escape rate
\begin{equation}
C_{0}=\frac{eB_{0}}{\omega_{0}m_{e}c}=1.76\times10^{6}\omega_{0}^{-1}\biggl(\frac{B_{0}}{0.1~\rm G}\biggr)~~\rm s^{-1}\,,
\label{Eq:28}
\end{equation}
and the initial \emph{Bohm} diffusive escape rate
\begin{eqnarray}
F_{0}&=&\frac{\eta_{0} m_{e}c^{3}}{r_{s}^{2}e B_{0}}\nonumber\\&=&5.12\times10^{-20}\eta_{0}\biggl(\frac{r_{s}}{10^{17}~\rm cm}\biggr)^{-2}\biggl(\frac{B_{0}}{0.1~\rm G}\biggr)^{-1}~~\rm s^{-1}\,.
\label{Eq:29}
\end{eqnarray}

\begin{figure}
	\centering
		\includegraphics[height=6cm,width=8cm]{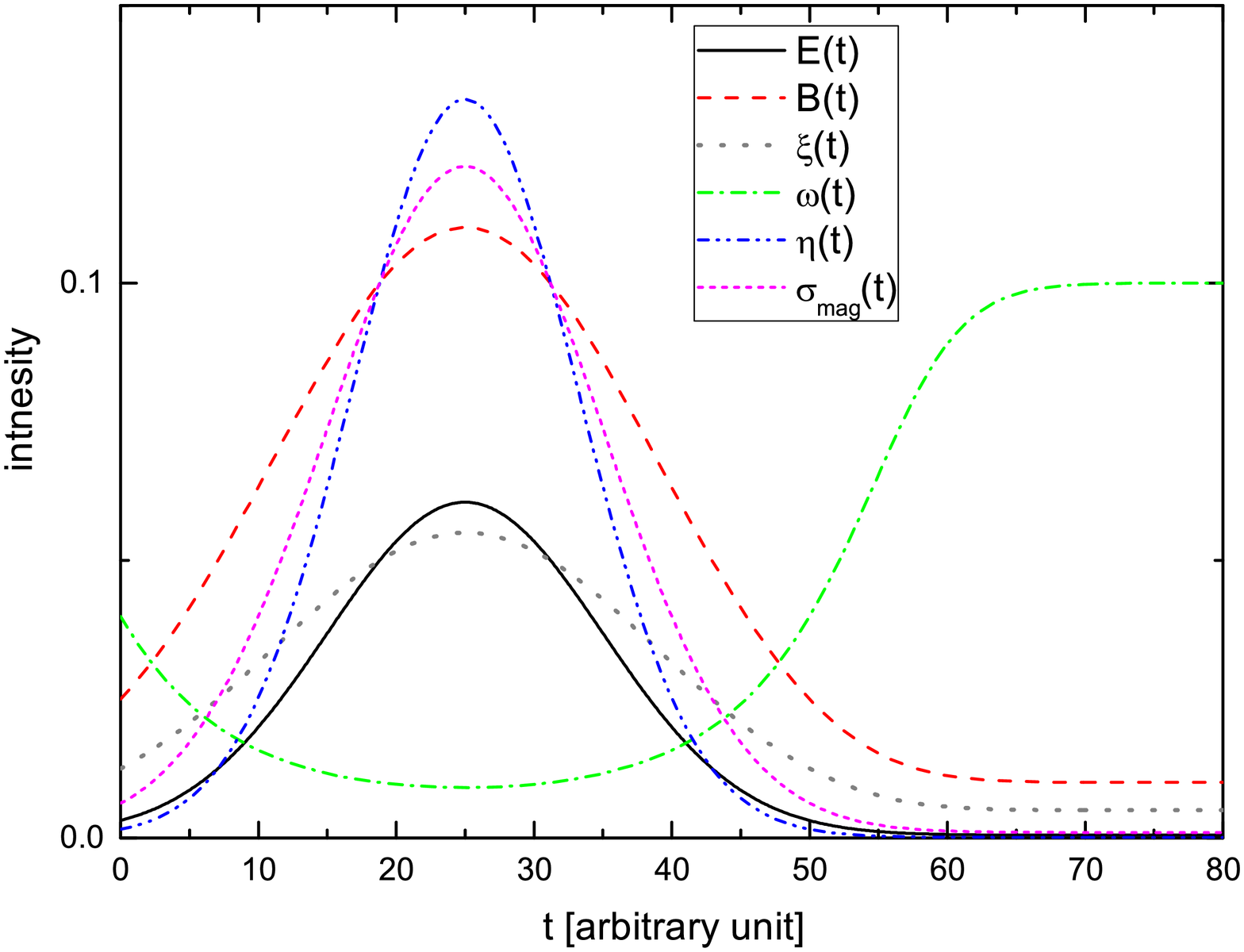}
	\caption{An example of the time-dependent parameters for a given time profile function. In this sample, we assume the initial values of the model parameters are as follows: $E_{0}=0.0005~\rm G$, $B_{0}=0.01~\rm G$, $\xi_{0}=0.005~\rm$, $\sigma_{\rm mag,0}=0.001$, $\eta_{0}=0.0001$, and $\omega_{0}=0.1$. We adopt a symmetrical time profile with $t_{*}=25$, $\alpha=120.0$, $\theta_{1}=10.0$, and $\theta_{2}=10.0$.}
	\label{Fig:2}
\end{figure}

We issue that the shock-induced magnetic reconnection may result in exciting the time-varying EM fields. Since there are some direct or indirect correlation between the theory parameter and EM fields (e.g., \citealt{2018ApJ...853...16K,2019ApJ...872...65K,2019ApJ...873....7Z}), the variable EM fields could induce to change other physical parameters. The precise time dependence of physical parameters are not known, but we can infer qualitative temporal trends by careful inspection of the time domain data products of the flare. It is convenient to deduce the time-dependent electric field, $E(t)$, and magnetic field, $B(t)$, as
\begin{equation}
E(t)=E_{0}\phi(t)\,,~~~~B(t)=B_{0}\sqrt{\phi(t)}\,.
\label{Eq:30}
\end{equation}
Consequently, we obtain the time-dependent efficiency factor of shock acceleration, $\xi(t)=\xi_{0}\sqrt{\phi(t)}$, the time-dependent dimensionless parameter that accounts for time dilation and obliquity in the relativistic shock, $\omega(t)=\omega_{0}[\phi(t)]^{-1/2}$, the time-dependent dimensionless parameter to represent the mean-free path relative to the Larmor radius, $\eta(t)=\eta_{0}[\phi(t)]^{3/2}$, and the time-dependent magnetization parameter, $\sigma_{\rm mag}(t)=\sigma_{\rm mag,0}\phi(t)$. The time-dependent physical parameters for a given time profile function are plotted in Figure {\ref{Fig:2}}. We note that the figure presents an example of the time-dependent parameters. The variability of the parameters with time is perceptible.

Application of the model requires the specification of the theory parameters $B_{\rm 0}$, $E_{\rm 0}$, $\xi_{\rm 0}$, $\sigma_{\rm mag, 0}$, $\eta_{0}$, $\omega_{\rm 0}$, $\delta$, $r_{\rm s}$, $u_{\rm ex}$, $N_{\rm ini}$, and $Q_{\rm 0}$. In our approach, we set $u_{\rm ex}$ and $r_{\rm s}$ as constant in all of our numerical calculations. The value corresponds to the characteristic luminosity of the accretion disk surrounding the central supermassive black hole and the observed variability timescale, respectively. The model free parameters $B_{\rm 0}$, $E_{\rm 0}$, $\xi_{\rm 0}$, $\sigma_{\rm mag, 0}$, $\eta_{0}$, $\omega_{\rm 0}$, $\delta$, $N_{\rm ini}$, and $Q_{\rm 0}$ are varied until a reasonable qualitative fit to the SED is obtained for the pre-flare steady-state. Once the model parameter values for the pre-flare steady-state have been established, we can utilize the physical parameters selected above and consider resulting particle spectral as an initial condition, but we change the particle spectral parameters to $B(t)=B_{\rm 0}\sqrt{\phi(t)}$, $E(t)=E_{\rm 0}\phi(t)$, $\xi(t)=\xi_{\rm 0}\sqrt{\phi(t)}$, $\sigma_{\rm mag}(t)=\sigma_{\rm mag, 0}\phi(t)$, $\eta(t)=\eta_{0}[\phi(t)]^{3/2}$, $\omega(t)=\omega_{\rm 0}[\phi(t)]^{-1/2}$, and $Q(t)=Q_{\rm 0}\phi(t)$ to reproduce the SED during the flare-state and post-flare state. For the time-profile-function, $\phi(t)$, we treat $\alpha$, $\theta_{1}$, and $\theta_{2}$ as free parameters. Since the function $\phi(t)$ increases from the initial value $\phi(0)\simeq1$, and reaches a maximum value at the given time $t_{*}$, we can pick the time $t_{\rm peak}$ at peak of flare from the light curves, and we set $t_{*}=t_{\rm peak}$. In these scenarios, the model for the flare requires twelve free parameters ($B_{\rm 0}$, $E_{\rm 0}$, $\xi_{\rm 0}$, $\sigma_{\rm mag, 0}$, $\eta_{0}$, $\omega_{\rm 0}$, $\delta$, $N_{\rm ini}$, $Q_{\rm 0}$, $\alpha$, $\theta_{1}$, and $\theta_{2}$).

\section{Evolution of the particle distribution}
In several simple cases, we are able to find an analytical, time dependent solution to the particle transport equation (see e.g., \citealt{2018PASP..130h3001Z} for a review). However, in the case of the SSC emission, where the energetic particles are interacting with the synchrotron radiation field that has generated this radiation, the partial differential equation, Eq. (\ref{Eq:4}), becomes a partial differential integral equation that is difficult to solve analytically. Therefore, we use a numerical approach (e.g., \citealt{1999MNRAS.306..551C,2006A&A...453...47K,2011MNRAS.416.2368C,2011ApJ...728..105Z}) to find the time-dependent solutions. Throughout the paper, we define the system evolution timescales in units of $t_{\rm cr}$ with $t_{\rm cr}=r_{\rm s}/c$.

\subsection{Validation of the numerical code}
Using a steady-state Green' function resulting from the reprocessing of mono-energetic seed particles, the exact particle distribution is solved by \cite{2019ApJ...873....7Z} from a generalized transport equation. It is important to note that for the case of injecting mono-energetic particles, a stationary three-segment particle spectrum can be expected. For the purpose of justifying the numerical code, we compare the numerical solution with exact particle distribution in two tests. We assume continuous mono-energetic particles are injected into a region, and they suffer from the acceleration, cooling and escape processes. We evolve the particle distributions with time up to a steady state. A constant time profile function, $\phi(t)=1$, is chosen during the evolution of the system.

\begin{figure}
\centering
   \includegraphics[height=6cm,width=8cm]{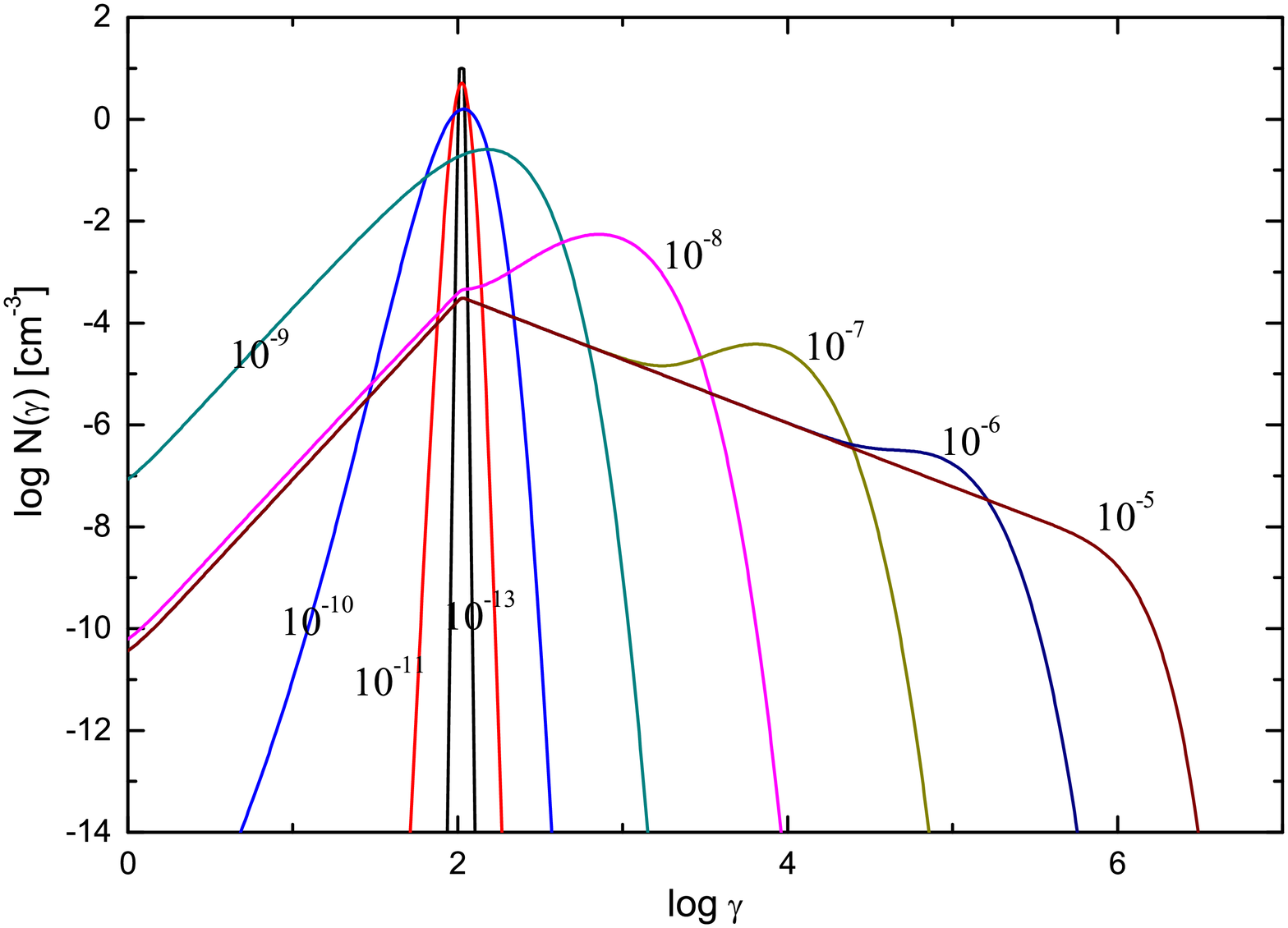}
  \includegraphics[height=6cm,width=8cm]{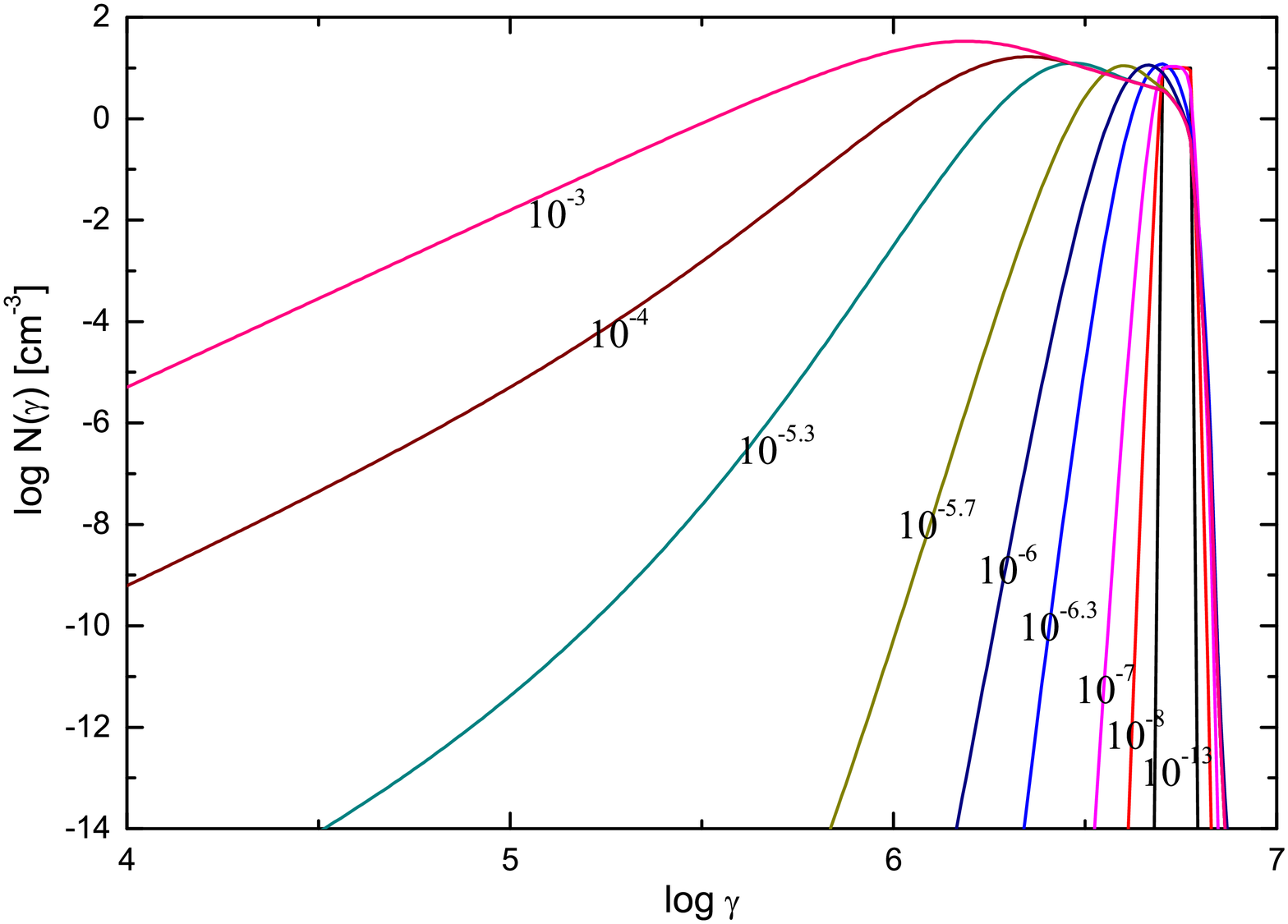}
\caption{Evolution of the particle distributions in the case of constant EM fields. The top panel shows the evolution of the initial spectrum $N_{\rm ini}(\gamma)=10~\rm cm^{-3}$ and the initial injected profile $Q_{0}(\gamma)=10~\rm cm^{-3}~s^{-1}$ with the particle Lorentz factor $1.0\times10^{2}\leq\gamma\leq1.1\times10^{2}$. The bottom panel shows the evolution of the initial spectrum $N_{\rm ini}(\gamma)=10~\rm cm^{-3}$ and the initial injected profile $Q_{0}(\gamma)=10~\rm cm^{-3}~s^{-1}$ with the particle Lorentz factor $5.0\times10^{6}\leq\gamma\leq6.0\times10^{6}$. We adopt the parameters of particle spectrum as follows: $B=0.01~\rm G$, $E/B=0.00001$, $\xi=0.1$, $\sigma_{\rm mag}=0.001$, $\eta=0.001$, $\omega=0.7$, $r_{\rm s}=1.0\times10^{16}~\rm cm$, $u_{\rm ex}=3.74~\rm erg~cm^{-3}$. Marks near the color curves represent the evolution timescales in units of $t_{\rm cr}$.}
  \label{Fig:3}
\end{figure}

The first test assumes a constant density of the particles, $N_{\rm ini}(\gamma)=10~\rm cm^{-3}$, in a very narrow initial distribution with $1.0\times10^{2}\leq\gamma\leq1.1\times10^{2}$. First, we calculate the evolution of the spectrum, where the particles with an initial profile $Q_{0}(\gamma)=10~\rm cm^{-3}~s^{-1}$ over the whole Lorentz factor range of the initial spectrum are continuously injected. In this simulation, the acceleration process attempts to increase the energy for all of the injected particles. The simulation presented in Figure \ref{Fig:3} (left panel) shows that the particle density systematically decreases in the injected Lorentz factor ranges while it simultaneous increases around the equilibrium Lorentz factor. Below the maximal injected Lorentz factor, the spectral slope evolves from an exponential to a power law. When a dynamic equilibrium is established, the competition between the acceleration and the escape produces a power law distribution that extends from the maximal Lorentz factor of the injected particles up to the equilibrium Lorentz factor. Above the equilibrium Lorentz factor, both the diffusion and cooling processes produce an exponential cut-off.

We next simulate the case of the same density and initial injection profile of the particles with the above test, but in a higher Lorentz factor distribution with $5.0\times10^{6}\leq\gamma\leq6.0\times10^{6}$. We show the evolution of this case in Figure \ref{Fig:3} (right panel). It can be seen that, instead of the mono-energetic distribution, we obtain a peaked spectrum that is more extended in the Lorentz factor range than the injected spectrum. The figure exhibits multiple regions of interest for discussion. At the beginning, the cooling process dominates the evolution of the particle spectrum. Therefore, the particle distribution moves towards the equilibrium Lorentz factor. When the peak Lorentz factor reaches the location of the equilibrium Lorentz factor, the shift in the peak ceases. After this, the part of distribution which located below the equilibrium Lorentz factor starts to change slope from an exponential to a power law. Above the equilibrium Lorentz factor, but below the minimal injected Lorentz factor, the evolution is dominated by the cooling and the escape that results in a power law distribution. Above the minimal injected Lorentz factor, the evolution is almost the same as the first test.

It can be seen that when the system evolution timescale exceeds $10^{-5}t_{\rm cr}$ in the case of low energetic particles injection and $10^{-3}t_{\rm cr}$ in the case of high energetic particles injection, the evolutions of the particle spectrum take the form of a stationary three-segment (or the double break power-law) particle distribution. Assuming a constant soft photon field, the stationary particle transport equation can be solved analytically  (\citealt{2019ApJ...873....7Z}). The analytical particle distribution shows the picture that a cusp center is surrounded by two power-law wings with an exponential high energy cutoff. We note that the shape of particle spectrum resulting from the numerical approach in the context is similar to it resulting from analytical approach. Since there is a lack of coherence the synchrotron radiation field with inverse Compton cooling in our early work (\citealt{2019ApJ...873....7Z}), we cannot expect a consistency between the numerical solution in this text and exact particle distribution.

\subsection{A simple evolution of the particle spectrum}
The influences of competing factors including the acceleration, injection, escape, and cooling of particles on the evolution of particle distributions have been exhibited in the previous two simulations. Here, we concentrate on the time-dependent changes in the physical parameters when the plasma blob is in contact with the shock in these two tests. In order to do this, we assume that both the initial distributions and the initial injection are equivalent to the profiles used in the previous tests, but we take the time dependent physical parameters into account during the evolution of the particle distributions. We adopt a symmetrical time profile with $\alpha=20.0$, $\theta_{1}=0.1~\rm s$, $\theta_{2}=0.1~\rm s$ and $t_{*}=10^{-6}t_{\rm cr}$ in following tests.

\begin{figure}
\centering
\includegraphics[height=6cm,width=8cm]{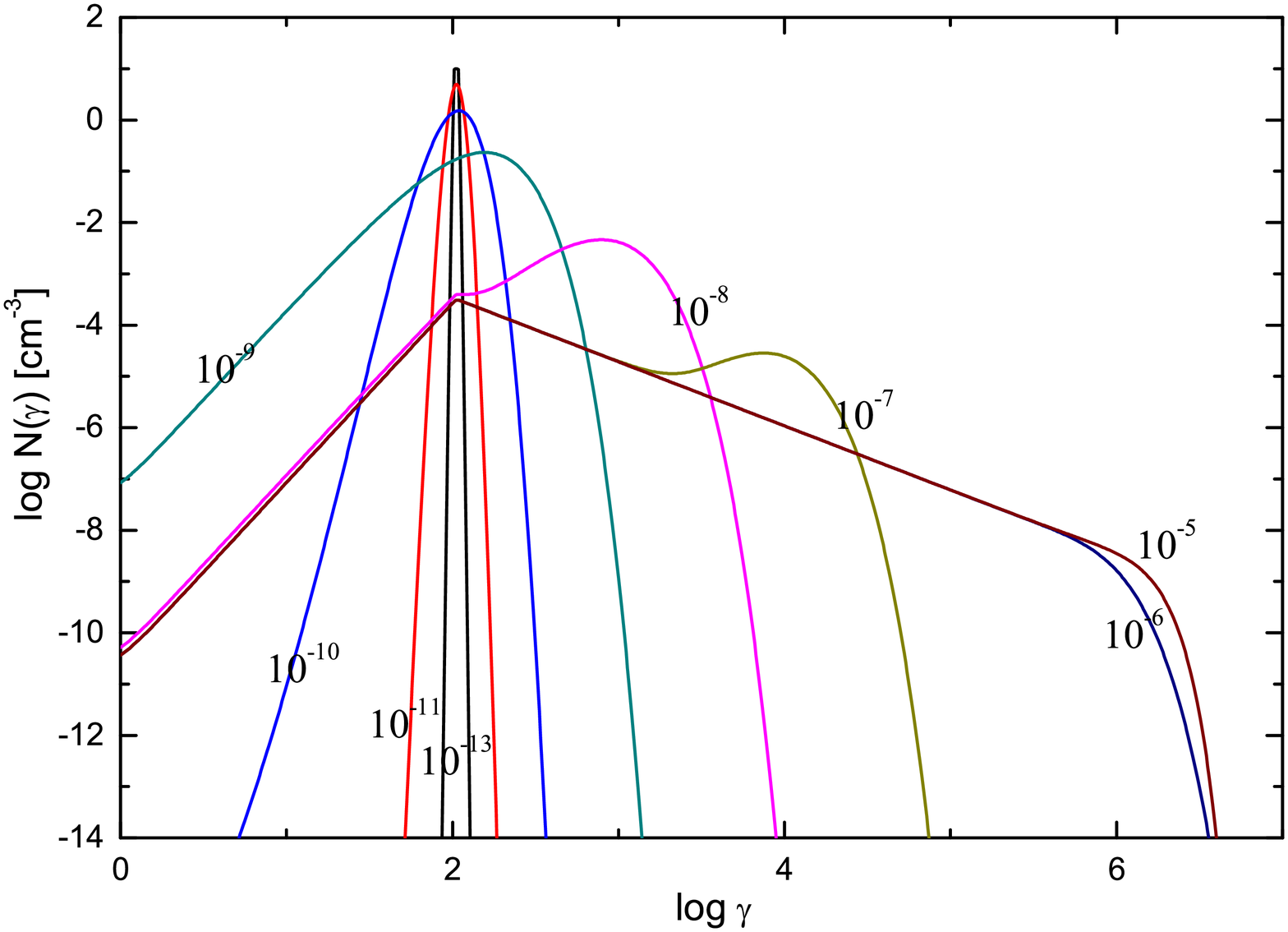}
  \includegraphics[height=6cm,width=8cm]{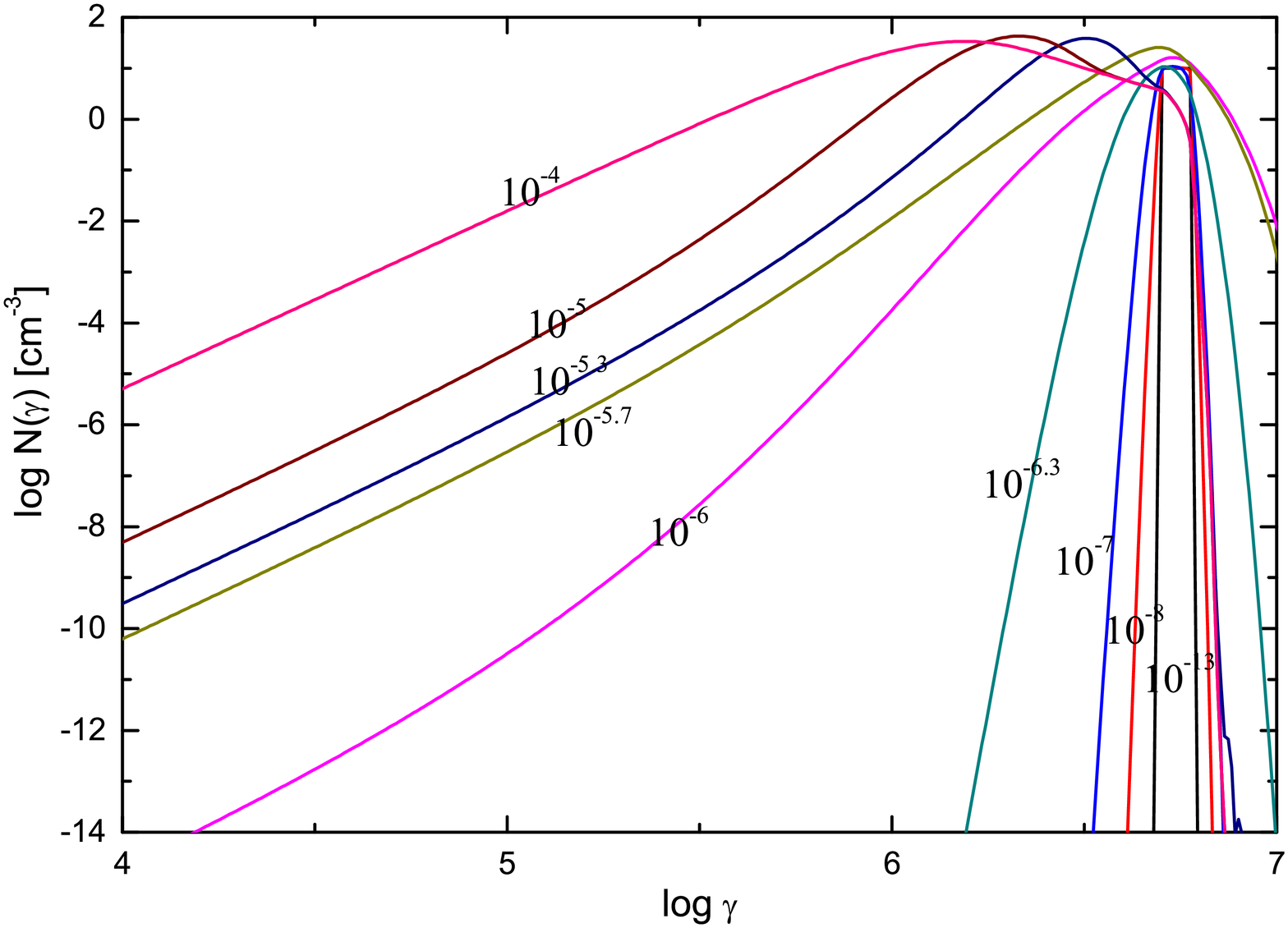}
\caption{Evolution of the particle distributions in the case of time-dependent EM fields. The top panel shows the evolution of the initial spectrum $N_{\rm ini}(\gamma)=10~\rm cm^{-3}$ and the initial injected profile $Q_{0}(\gamma)=10~\rm cm^{-3}~s^{-1}$ with the particle Lorentz factor $1.0\times10^{2}\leq\gamma\leq1.1\times10^{2}$. The bottom panel shows the evolution of the initial spectrum $N_{\rm ini}(\gamma)=10~\rm cm^{-3}$ and the initial injected profile $Q_{0}(\gamma)=10~\rm cm^{-3}~s^{-1}$ with the particle Lorentz factor $5.0\times10^{6}\leq\gamma\leq6.0\times10^{6}$. We adopt the parameters of particle spectrum as follows: $B_{0}=0.01~\rm G$, $E_{0}/B_{0}=0.00001$, $\xi_{0}=0.1$, $\sigma_{\rm mag,0}=0.001$, $\eta_{0}=0.001$, $\omega_{0}=0.7$, $r_{\rm s}=1.0\times10^{16}~\rm cm$, $u_{\rm ex}=3.74~\rm erg~cm^{-3}$. We assume a symmetrical time profile with $\alpha=20.0$, $\theta_{1}=0.1~\rm s$, $\theta_{2}=0.1~\rm s$ and $t_{*}=10^{-6}t_{\rm cr}$. Marks near the color curves represent the evolution timescales in units of $t_{\rm cr}$ with $t_{\rm cr}=r_{s}/c$.}
\label{Fig:4}
\end{figure}

The evolutions of the electron distributions for each profiles (initial distribution and continuous injection) are plotted in Figure {\ref{Fig:4}}. Being analogous to the evolution in the previous test, when the system evolution timescale exceeds $10^{-5}t_{\rm cr}$ in the case of low energetic particles injection and $10^{-4}t_{\rm cr}$ in the case of high energetic particles injection, the evolutions of the particle spectrum take the form of a stationary three-segment particle distribution. By comparing Figures 3 and 4, we note that the particle evolution appears very similar up to the time when the system evolutions are close to the given time, $t_{*}$, and the shape of the particle distribution is transformed distinctly. On the other hand, a transient fluctuation of the EM fields can boost the acceleration efficiency, and result in particle distribution extensions to higher Lorentz factor regions.

It is believed that both the shock acceleration and the stochastic acceleration are difficult to trigger in the very low energy of the initial or injected particles. Therefore, some pre-acceleration of the particles is required in order to make the shock or stochastic acceleration operate efficiently. Since the electrostatic acceleration is also efficient in the vicinity of the shock front, such pre-acceleration process in the context can occur and result in higher energy for the initial and/or injected particles.
\subsection{Equilibrium timescale of the system evolution}
The transport equation, Eq. (\ref{Eq:2}), in the current context takes the shock, electrostatic, stochastic acceleration, synchrotron and IC scattering loss, as well as the particle injection and escape into account. If we neglect the particle injection and escape, it is convenient to obtain the theoretical equilibrium momentum for combination of stochastic acceleration and momentum loss by balancing the stochastic momentum gain rate, $\dot{p}_{\rm stoch}$, with the momentum-loss rate, $\dot{p}_{\rm loss}$, so that the equation
\begin{equation}
(\dot{p}_{\rm stoch}+\dot{p}_{\rm loss})|_{p=p_{\rm e, stoch}}=0\;,
\label{Eq:31}
\end{equation}
can be established. The equilibrium Lorentz factor for stochastic acceleration versus momentum loss is therefore given by
\begin{equation}
\gamma_{\rm e, stoch}=\frac{p_{\rm e, stoch}}{m_{e}c}=\sqrt{\frac{9D_{0}m_{e}c}{4\sigma_{T}u_{\rm ph}}}\;.
\label{Eq:32}
\end{equation}
With the same approach, we can establish the following equations
\begin{equation}
(\dot{p}_{\rm sh}+\dot{p}_{\rm loss})|_{p=p_{\rm e, sh}}=0\;,
\label{Eq:33}
\end{equation}
and
\begin{equation}
(\dot{p}_{\rm elec}+\dot{p}_{\rm loss})|_{p=p_{\rm e, elec}}=0\;,
\label{Eq:34}
\end{equation}
which yields the equilibrium Lorentz factor for shock acceleration versus momentum loss,
\begin{equation}
\gamma_{\rm e, sh}=\frac{p_{\rm e, sh}}{m_{e}c}=\sqrt{\frac{9\xi_{0} eB_{0}}{16\sigma_{T}u_{\rm ph}}}\;,
\label{Eq:35}
\end{equation}
and the equilibrium Lorentz factor for electrostatic acceleration versus momentum loss,
\begin{equation}
\gamma_{\rm e, elec}=\frac{p_{\rm e, elec}}{m_{e}c}=\sqrt{\frac{3eE_{0}}{4\sigma_{T}u_{\rm ph}}}\;,
\label{Eq:36}
\end{equation}
respectively. By combining the acceleration rate of particles, we can estimate the equilibrium timescale, $t_{\rm e, stoch}=\gamma_{\rm e, stoch}/(3D_{0})$, $t_{\rm e, sh}=\gamma_{\rm e, sh}/\dot{\gamma}_{\rm sh,0}$, and $t_{\rm e, elec}=\gamma_{\rm e, elec}/\dot{\gamma}_{\rm elec,0}$, respectively. We can expect a characteristic timescale that it takes the system evolution achievement to a steady state, $t_{\rm e}=\rm min~[t_{\rm e, stoch},t_{\rm e, sh},t_{\rm e, elec}]$. If we take the values of the particle parameters adopted in the above tests into account, and we assume $u_{\rm syn}=10u_{\rm B}$, we can obtain $t_{\rm e, stoch}=6.8~\rm s$, $t_{\rm e, sh}=24.9~\rm s$, and $t_{\rm e, elec}=2162.0~\rm s$. One can find that characteristic timescale is in agreement with the results of the numerical approaches.

In the context of particle evolution approach, we should compare the characteristic timescale, $t_{\rm e}$, where it takes the system evolution achievement to a steady state, with the adopted system evolution timescale or the observed time bin or the typical variability timescale of blazars, $t_{\rm var}$. The timescales in the observer's frame are corrected by Doppler factor and redshift. If it satisfies $t_{\rm e}\leq t_{\rm var}$, the steady state particle distribution has been shaped before the evolution time elapsed. In this point, the impress of acceleration and cooling are hard to be distinguished from the evolution on the particle distribution for the given theory parameters. That is, in this case, we can consider a steady state particle distribution for the given theory parameters. Since we assume the time dependent theory parameters in the model, we can find different steady state particle distribution with different values of theory parameters at the time. On the contrary, if it satisfies $t_{\rm e}\geq t_{\rm var}$, the acceleration and cooling are significant to the evolution on the particle distribution. In this case, the energy dependent time lag could be expected during the flare.

We note that the timescale of the evolution of the particle spectra to achieve a stationary state is the order of seconds to minutes. This is less than the observed time bin and/or the typical variability timescale of blazars. In this scenario, we consider the dynamic equilibrium of the system can be generated instantaneously by a competition between the acceleration and the cooling and/or escape of particles from the shock region. We expect the light curve profiles of blazars to be consistent with the particle spectra variability results from time-varying EM fields, rather than the effect of both the acceleration and the cooling processes.


\section{Expected Variability}
Our goal is to produce an integrated model that can simultaneously account for the variability of the energy spectrum and the light curve profile. Hence, we analyze the synchrotron and SSC emission generated by a homogeneous, spherical blob filled by electrons and magnetic fields. The particles are accelerated uniformly throughout the dissipated region. It is believed that for a given variability, all of the physical parameters play important roles for the steady state of pre-flaring, whereas the parameters of the time-profile-function are important for fitting the spectral evolution and light curves. On the theoretical side, shock and reconnection both can lead to magnetic field evolution but require very different magnetization. One can seen that a high magnetization results in a hard particle spectrum (\citealt{2019ApJ...873....7Z}). Since the spectral index of particle spectrum is too hard to fit the multi-wavelength SED of blazars (e.g., \citealt{2002ApJ...564...92C}), we suggest the model applies better in weak magnetization.

To apply the proposed theoretical model to the computation of the emission variability, we must input an initial particle distribution for the stable spectral energy distribution (SED) of pre-flaring. The previous tests exhibit that a stationary solution can be expected from the evolution of the transport equation. Therefore, we can consider the resulting steady-state spectrum as an initial particle distribution, but we solve the time-dependent electron distribution functions with the time-dependent EM fields. In this section, we present the expected variability for the solutions applicable to the synchrotron and SSC emission. Since the typical timescale of rapid TeV $\gamma$-ray flaring is in the magnitude of a few minutes (e.g., \citealt{2007ApJ...664L..71A,2007ApJ...669..862A}), in order to coincide the variability of theoretical photon spectra with observations, we adopt a given time $t_{*}=300~\rm s$ for the time profile function, and we consider a symmetrical time profile with $\alpha=1.2$, $\theta_{1}=100~\rm s$ and $\theta_{2}=100~\rm s$ . For the sake of simplicity we calculate only the case of a relatively low energy for the injected particles with a Lorentz factor, $1.0\times10^{2}\le\gamma\le1.1\times10^{2}$.

\begin{figure}
	\centering
		\includegraphics[height=6cm,width=8cm]{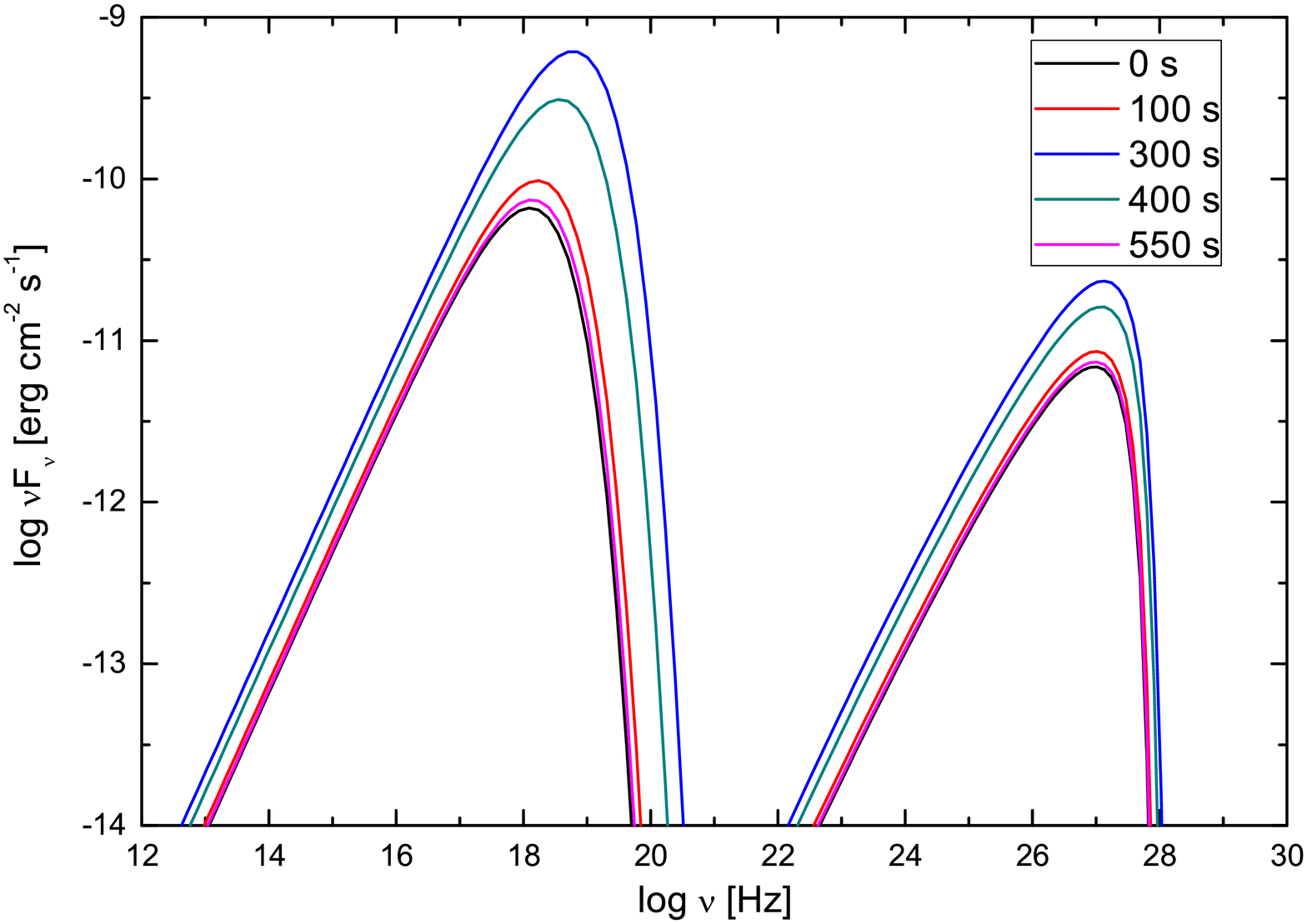}	
	     \includegraphics[height=6cm,width=8cm]{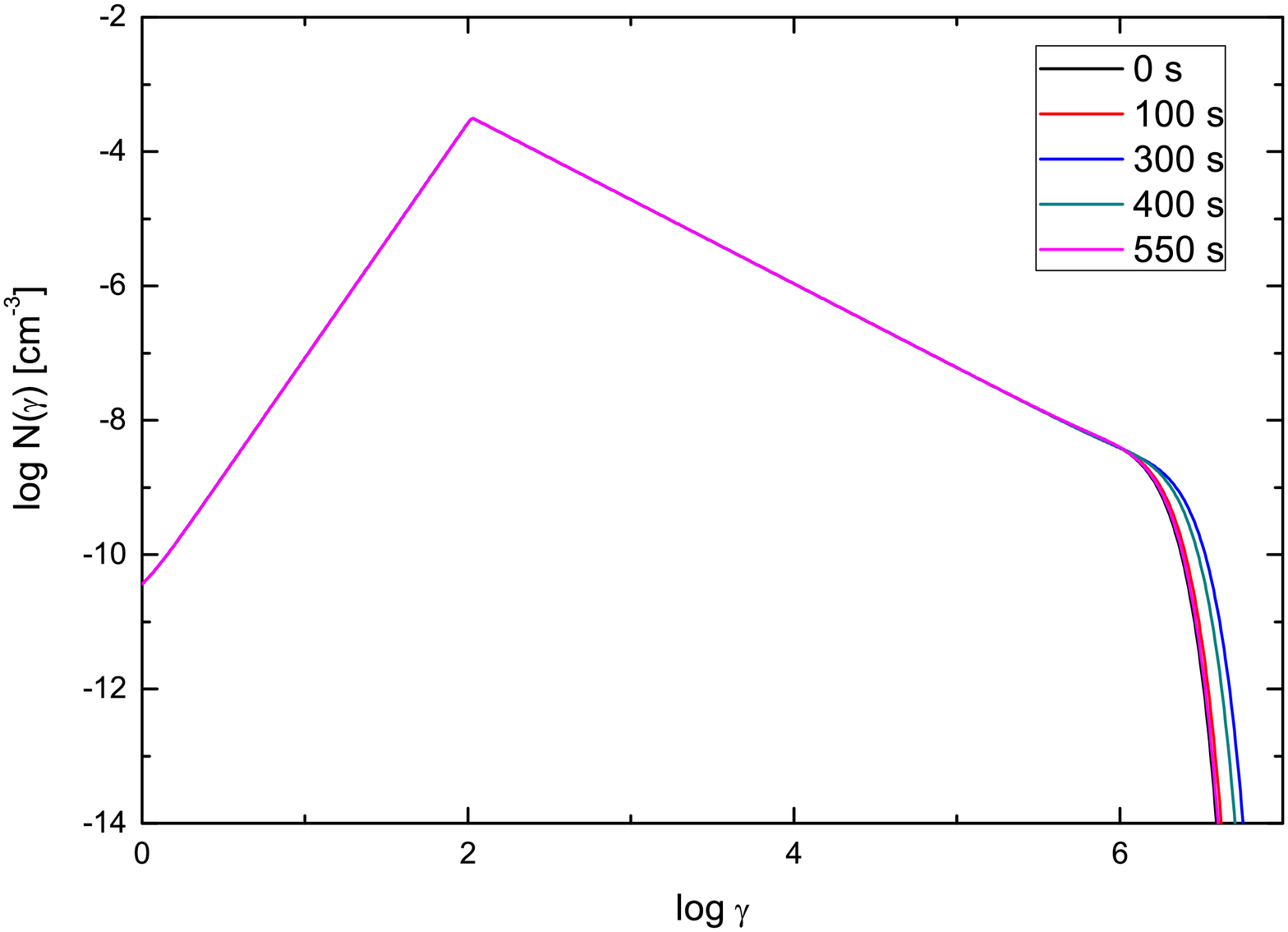}	
	\caption{The evolution of the theoretical photon spectra cased by a transient fluctuation of the EM fields (top panel). The associated electron distributions are also exhibited (bottom panel). We assume the initial value of the model parameters as follows: $B_{0}=0.01~\rm G$, $E_{0}/B_{0}=0.00001$, $\xi_{0}=0.1$, $\sigma_{\rm mag,0}=0.001$, $\eta_{0}=0.001$, $\omega_{0}=0.7$, $r_{\rm s}=1.0\times10^{16}~\rm cm$, $u_{\rm ex}=3.74~\rm erg~cm^{-3}$. We evolve the transport equation based on the initial value of the model parameters to a steady-state and input the stationary solution as a initial particle distribution. We consider a symmetrical time profile with $t_{*}=300~\rm s$, $\alpha=1.2$, $\theta_{1}=100~\rm s$, and $\theta_{1}=100~\rm s$. The particle spectrum is calculated at each point in the time from 0 s to 1000 s. Then, the theoretical photon spectra can be expected by both the synchrotron and SSC emission with Doppler factor $\delta=20$ and a absorption optical depth (e.g., \citealt{2011ApJ...728..105Z,2013ApJ...764..113Z,2018MNRAS.478.3855Z,2019ApJ...873....7Z}).}
	\label{Fig:5}
\end{figure}

The theoretical photon spectra plotted in Figure {\ref{Fig:5}} have been computed by assuming the time-dependent EM fields at each point in time. It is clear from Figure {\ref{Fig:5}} that the peak amplitude of the synchrotron emission is has an increasing trend and is moving toward higher frequencies; the peak position of the self-Compton emission is not as strong as the synchrotron emission. The very high energy $\gamma$-rays are significantly absorbed due to interactions with the extragalactic background light (EBL, e.g., \citealt{1967PhRv..155.1408G,1992ApJ...390L..49S}).
As a result, the absorption produces a cut-off in the spectra above the frequency $\nu_{\rm ic}\sim 10^{28}~\rm Hz$. Up to the time when the system evolution overwhelms the given time $t_{*}$, the peak of both the synchrotron and self-Compton emissions decrease in amplitude. The corresponding electron distributions for the theoretical photon spectra at different evolution timescale are also plotted in Figure {\ref{Fig:5}}. The figure relates to the fact that, if the time, at which the system evolution attains to a dynamic equilibrium state, less than the timescale of variabilities, we expect that the distribution function at each timescale closely resembles the stationary solution of the transport equation Eq. (\ref{Eq:4}). On the high-energy side, the particle spectrum is in good agreement with our previous tests, it extends to higher Lorentz factor regions.

\begin{figure}
	\centering
		\includegraphics[height=6cm,width=8cm]{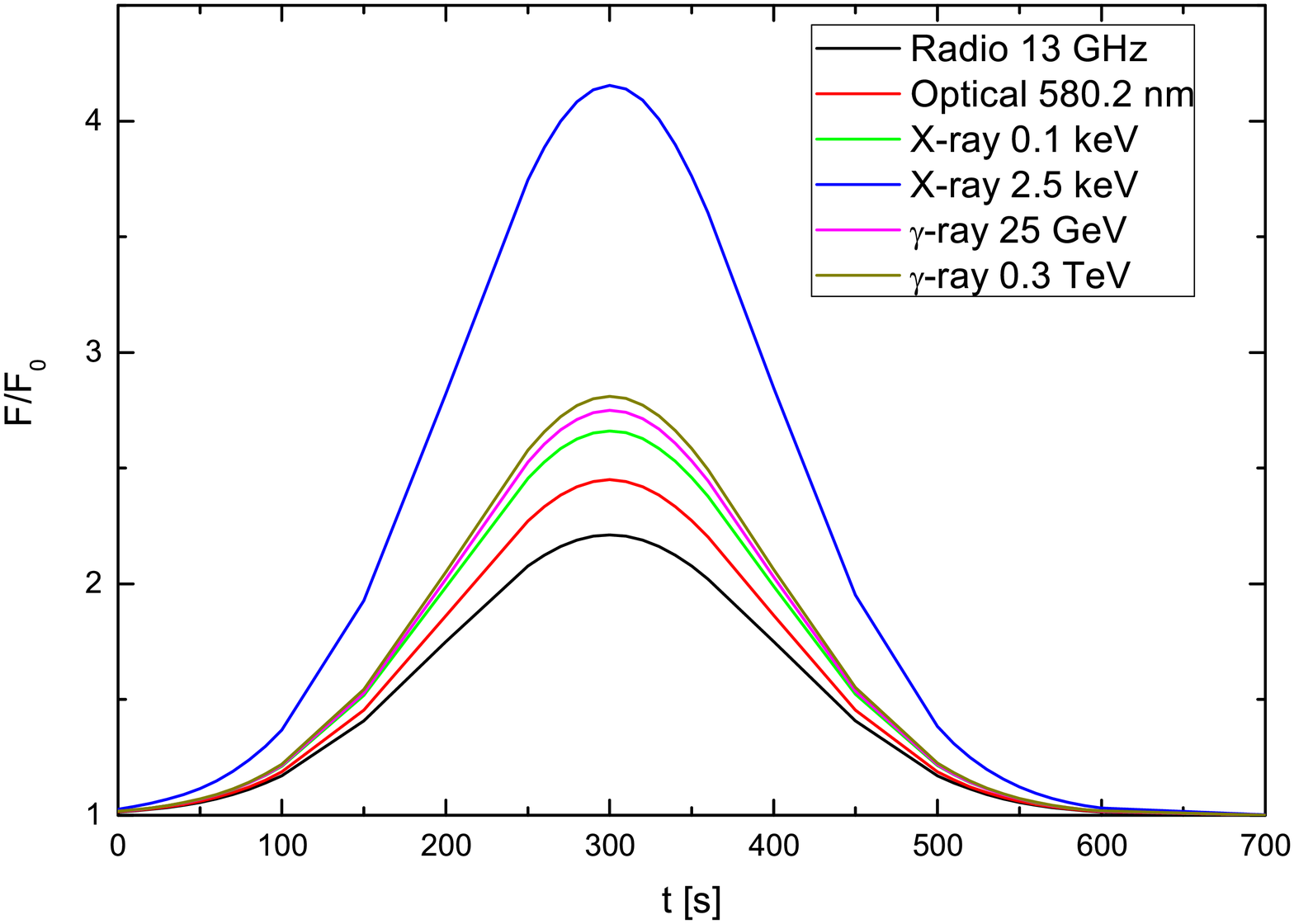}	
	\caption{The model expected multi-wavelength variabilities cased by a transient fluctuation of the EM fields. The symmetrical variability curves are able to roughly reproduce.}
	\label{Fig:6}
\end{figure}

The model's expected multi-wavelength variabilities are plotted in Figure {\ref{Fig:6}}. It can be seen that, from radio bands to high-energy $\gamma$-ray bands, the symmetrical light curves are reproduced. As we expected, a transient fluctuation of the EM fields can boost the acceleration efficiency that result in the strong X-ray and $\gamma$-ray flaring.

\begin{figure*}
	\begin{center}
    \begin{tabular}{c}
    		\includegraphics[height=10cm,width=13cm]{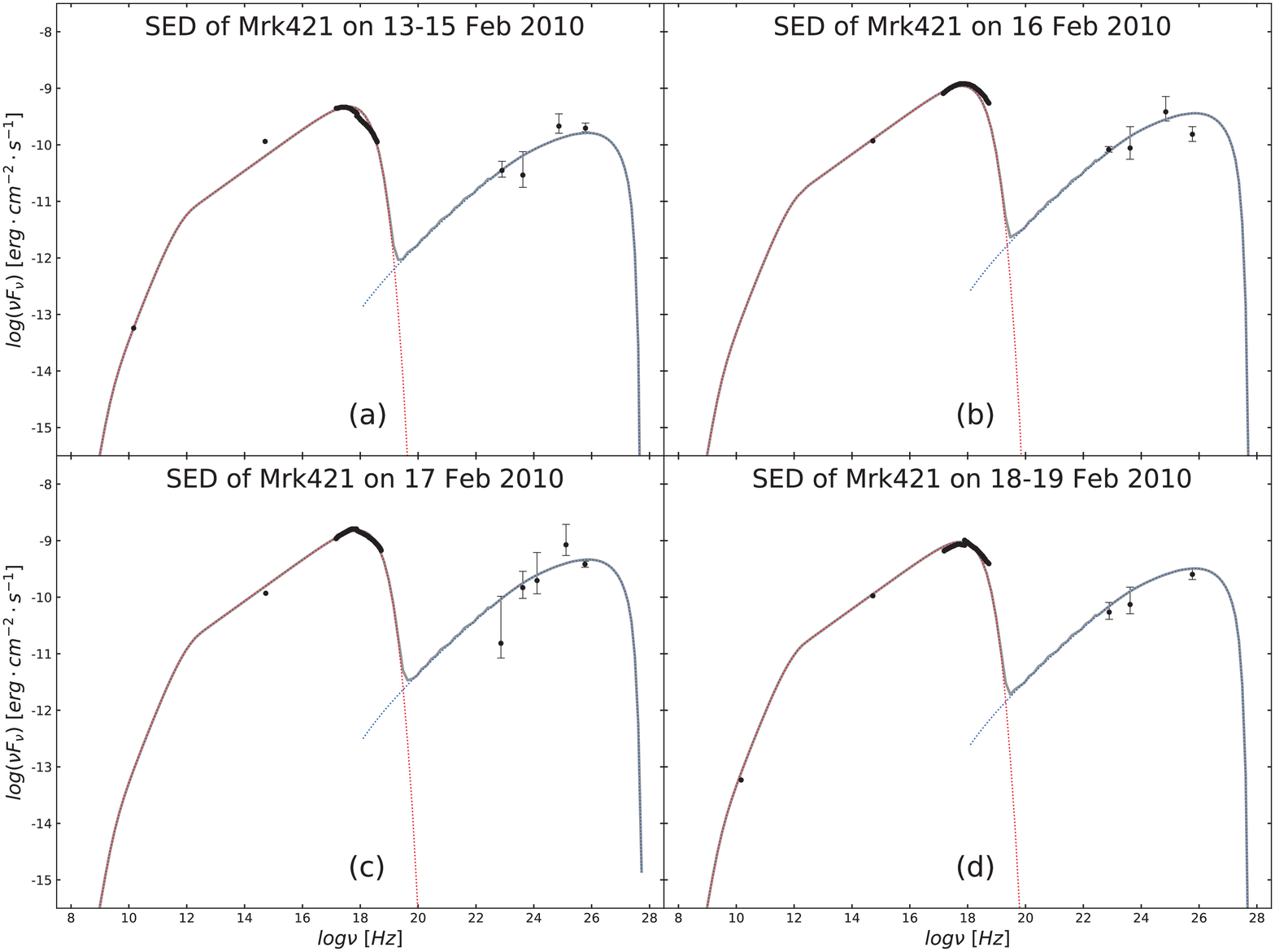}
	\end{tabular}
    \end{center}
       	
	\caption{Comparisons of model spectra with observed data for the BL Lac object Mrk 421. The solid circles show the SED during a giant flare on February 13-19, 2010 (\citealt{2012A&A...541A.140S}). Panel (a) shows the pre-flare state that is averaged over the observations taken during February 13-15, 2010; Panel (b) shows the X-ray and $\gamma$-ray flare state on February 16, 2010; Panel (c) shows the TeV $\gamma$-ray flare state on February 17, 2010; Panel (d) shows the post-flare state that is averaged over the observations taken during February 18-19, 2010. The theoretical photon spectra for each state is shown as solid curve. }
	\label{Fig:7}
\end{figure*}

\begin{figure*}
	\begin{center}
     \begin{tabular}{c}
		\includegraphics[height=22.5cm,width=15cm]{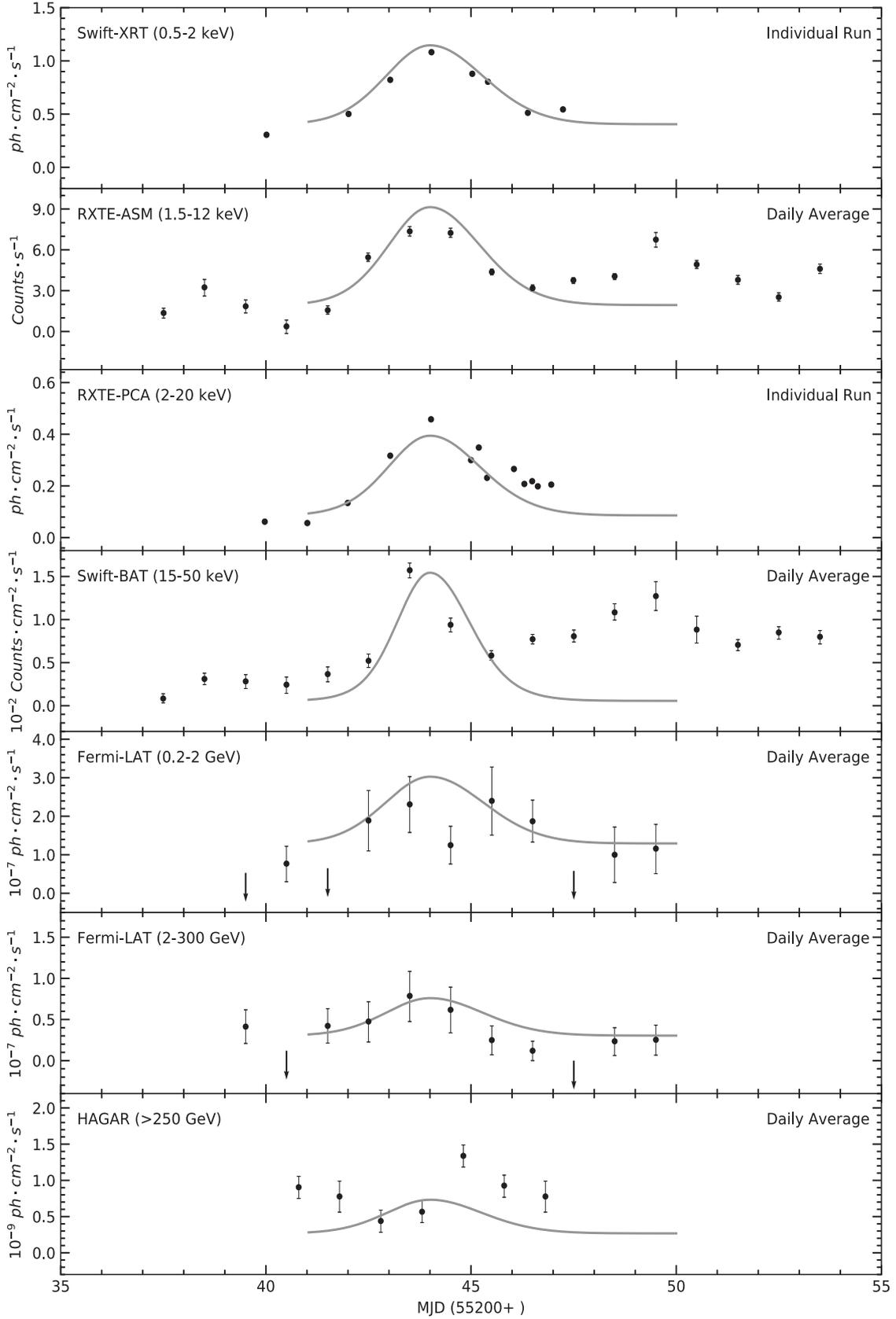}
    \end{tabular}
    \end{center}
	\caption{Multi-wavelength light curve of Mrk 421 during February 10 - 26, 2010. The separated in different energy bands from top to bottom, Swift-XRT (0.5-2 keV), RXTE-ASM (1.5-12 keV), RXTE-PCA (2-20 keV), Swift-BAT (15-50 keV), Fermi-LAT (0.2-2 GeV), Fermi-LAT (2-300 GeV), HAGAR (>250 GeV). The solid curves show a model expected variabilities. We set $t=0$ at the time of 12h on February 14, 2010.}
	\label{Fig:8}
\end{figure*}

As mentioned above, in order to check whether the discussed scenario is able to explain the variability of the energy spectrum and the light curve, we apply the results of the simulation for a giant flare of Mrk 421 observed by High Altitude GAmma Ray (HAGAR) telescope, \emph{Fermi}-LAT, RXTE-ASM, Swift-BAT, Swift-XRT, RXTE-PCA (\citealt{2012A&A...541A.140S}) in February, 2010. In Figure \ref{Fig:7}, we compare the model's generated spectra with the observed data for the BL Lac object Mrk 421. The model parameters are varied in order to obtain a reasonably good qualitative fit to pre-flaring data set. We list the value of the parameters in Table \ref{tab1}. It is clear from Figure \ref{Fig:7} that the electron transport model considered here is able to roughly reproduce the observed spectra for each of the observed state. Once the corresponding electron distribution for the pre-flaring state is established, we input the electron distribution as an initial condition. This reproduces the variability of the energy spectrum and the light curve with an impulse variability for the EM fields. As shown in Figure {\ref{Fig:7}} and {\ref{Fig:8}}, the model in the context reproduces the flaring spectral and light curve well if we select the parameters of the time profile function in the comoving frame, $t_{*}=8.01\times10^{6}~\rm s$, $\alpha=0.67$, $\theta_{1}=3.1\times10^{6}~\rm s$, and $\theta_{2}=3.72\times10^{6}~\rm s$. It is worth noting that, when we perform the fitting, these timescales are transformed to the observed frame by $t_{\rm obs}=(1+z)t_{\rm in}/\delta$.

\begin{table*}
\centering
\caption{Physical parameters of the model spectra}\label{tab1}
\begin{tabular}{lclc}
\hline \hline
\multirow{2}{*}{Parameters} &
\multirow{2}{*}{Pre-flare} &
\multirow{2}{*}{Parameters$^{*}$} &
X-ray and $\gamma$-ray flare, TeV flare, post-flare\\
   \cline{4-4}
   &&&$t_{*}=8.01\times10^{6}s,\alpha=0.67,\theta_{1}=3.1\times10^{6}s,\theta_{2}=3.72\times10^{6}s$\\
\hline
$B_{0}~[\rm G]$                    & 0.02 &                            $B(t)~[\rm G]$     &$B_{0}\sqrt{\phi(t)}$       \\
 $E_{0}/B_{0}$                     & 0.34 &                            $E(t)~[\rm G]$     &$E_{0}\phi(t)$              \\
 $\xi_{0}$                         & 0.03                             & $\xi(t)$   &    $\xi_{0}\sqrt{\phi(t)}$                     \\
 $\sigma_{\rm mag, 0}$             & 0.001                            & $\sigma_{\rm mag}(t)$&    $\sigma_{\rm mag,0}\phi(t)$                       \\
 $\eta_{0}$                        & 0.0069                           & $\eta(t)$&       $\eta_{0}[\phi(t)]^{3/2}$                   \\
 $\omega_{0}$                      & 0.8                              & $\omega(t)$&$\omega_{0}[\phi(t)]^{-1/2}$                             \\
 $\delta$                          & 32                               & $\delta$&32                              \\
 $r_{s}~[\rm cm]$                  & $2.3\times10^{16}$               & $r_{s}~[\rm cm]$                  & $2.3\times10^{16}$              \\
 $u_{\rm ex}~[\rm erg~cm^{-3}]$    & 14.92                            & $u_{\rm ex}~[\rm erg~cm^{-3}]$    & 14.92                           \\
 $N_{\rm ini}~[\rm cm^{-3}]$       & 50~in $460\le\gamma\le490$       & $N_{\rm ini}~[\rm cm^{-3}]$& the induced particle spectra from pre-flaring state     \\
 $Q_{\rm 0}~[\rm cm^{-3}~s^{-1}]$  & 55~in $460\le\gamma\le490$       & $Q(t)~[\rm cm^{-3}~s^{-1}]$&$Q_{\rm 0}\phi(t)$     \\
\hline
\end{tabular}\\
{Note: $^{1}$ A time profile of positive skewness is adopted.}
\end{table*}

\section{Conclusion and Discussion}
An exclusive emission characteristic for the jet of blazars is that there is a noticeable and rapid flux variability at all observed frequencies. Some modeling efforts have been made in explaining blazar flares. For example, \cite{2011MNRAS.416.2368C} employ a time-dependent two component radiative transfer model where the light travel-time effects are fully considered. Their model suggests that the variability is produced by the injection of relativistic electrons when a shock front crosses the emission region (\citealt{2011MNRAS.416.2368C}). Alternatively, the observed large polarization angle swings favors a scenario in which magnetic energy dissipation is the primary driver of the flare event (\citealt{2015ApJ...804...58Z}). The exact solution for the Fourier transform of the particle distribution is presented in \cite{2016ApJ...824..108L}; the results yield a complete first-principles physical explanation for both the formation of the observed time lags and the shape of the peak flare X-ray spectrum. This paper proposes a time-dependent particle evolution model to explain the variability of blazars. The model introduces time-dependent EM fields into the transport equation, supporting other physical quantities' variability with time. It allows us to understand how the impulse variability of EM fields affect the evolution of the particle distributions. The numerical simulation of evolution on the particle distributions indicate that the light curve profiles of blazars are consistent with the particle spectra variability results from time-varying EM fields, rather than the effects of the acceleration and cooling processes. Furthermore, we apply the model to both the energy spectra and multi-wavelength activity of BL Lac object Mrk 421. The results strongly indicate that the magnetic field evolution in the dissipated region of a blazar jet can account for the variability.

It can be seen that in our early work (\citealt{2020arXiv200906152Z}), if we focus on varying the magnetic field strength, since the dynamic equilibrium of the system reestablishes, we can find a continuous variability covering energies from high to low. Alternatively, if we focus on varying the electric field strength, the efficiency acceleration contains the electron populations around on the equilibrium energy, we can expect the shapes of these SED bumps to change. In addition, the flux variability is remarkable and the change of spectral slopes is significant at the X-ray and GeV-TeV $\gamma$-ray bands; however, as a result of that the low energy electrons are dominated by the electrostatic acceleration process, the variability is minor or not significant at the low frequency end of synchrotron component. As for the flaring timescales and optical polarization, they may be relate to the essential process of reconnection, we leave them in the future work.

The current evolution on the electron distribution is solved numerically from a generalized transport equation that contains the terms describing the electrostatic, first-order and second-order \emph{Fermi} acceleration, escape of particles due to both advection and spatial diffusion, in addition to energy losses due to the synchrotron emission and inverse-Compton scattering of both synchrotron and external ambient photon fields. It is interesting to find that the system needs a little time to accumulate a significant number of the particles around the equilibrium Lorentz, where the energy acquired by the particles in the acceleration process is radiated away. This scenario reveals that in the current approach, the evolution timescales of a system achieving a stationary state are significantly less than the earlier modeling efforts, where there is an absence of terms describing the electrostatic, first-order \emph{Fermi} acceleration, and the escape of particles due to advection, but assume a large acceleration timescales (\citealt{2006A&A...453...47K,2011ApJ...728..105Z}). So instantaneous system evolutions are likely to shield the brand of particle acceleration and cooling, such that the resulting variability is marked by an energy-dependent hard or soft lag.

In order to obtain a suitable particle distribution for modeling the SED of blazar, the present model takes into account an external ambient photon field as a free parameter. It is convenient to estimate the maximum energy of the synchrotron emission component peak in the case of neglecting the injection and the escape of particle populations,
\begin{eqnarray}
E_{\rm syn, peak}^{\rm in}(\gamma_{\rm max})&=&\gamma_{\rm max}^{2}m_{e}c^{2}\frac{B_{0}}{B_{\rm crit}}\nonumber\\
&\le&6.50~\rm MeV\biggl(\frac{B_{0}}{0.1~\rm G}\biggr)^{2}\biggl(\frac{u_{B_{0}}+u_{\rm ph}}{0.01~\rm erg~cm^{-3}}\biggr)^{-1}\nonumber\\&\times&\biggl(1+\frac{E_{0}}{B_{0}}\biggr)\;.
\label{Eq:37}
\end{eqnarray}
Here, the parameter $D_{\rm 0, max}$ satisfies the relationship, $D_{\rm 0, max}=eB_{0}/3m_{e}c$. While clear indications or evidence for the synchrotron peak actually reaching the MeV range (\citealt{2011A&A...534A.130K,2011MNRAS.414.3566T,2014ApJ...787..155T,2017A&A...598A..17C}) are still under debate, we can expect that there are a extreme high-energy synchrotron peaked sources (\citealt{2018MNRAS.477.4257C,2018MNRAS.480.2165A}) with $E_{\rm syn, peak}^{\rm obs}\simeq1~\rm keV$. If we adopt the energy of synchrotron peak $E_{\rm syn, peak}^{\rm obs}\simeq10~\rm keV$ with $\delta\sim10$, the value for the energy density of an ambient photon field is obtained by substituting $B_{0}$, $D_{0}$, $\xi_{0}$, and $E_{0}/B_{0}$ for the averaged case in Table 1 into Eq. (\ref{Eq:37}), which yields $u_{\rm ph}=31.21~\rm erg~cm^{-3}$. This value is three orders of magnitude above the energy density of a synchrotron photon field. Distinctly, the expected energy density of an ambient photon field is significantly larger than the external component included in the model to overcome the extreme parameters in the BL Lac object (e.g., \citealt{2008ApJ...684L..73A,2011ApJ...726...43A,2014ApJ...783...83L,2016MNRAS.461.1862K}). In these scenarios, we argue that the additional ambient photon field is required to equilibrate the efficient acceleration in a suitable energy regime for reproducing the synchrotron and SSC emission; nevertheless, it is considered as an external photon field in the context model. We do not propose an explanation to describe where the additional ambient photon field originates. However, it is interesting to speculate that this may be a result of the plasmoid density growing rapidly through mergers before leaving the reconnection region, i.e., forming ``monster'' plasmoids (\citealt{2010PhRvL.105w5002U}). Consequently, the deterministic cooling increases the energy density of the ambient photon field as an avalanche occurs. Since the change of magnetic field generally does not affect external Compton (EC) component in the context, for EC dominated flat spectrum radio quasars (FSRQs) we can expect orphan optical flares. A consequence of a flaring event being triggered by the magnetic field evolution, is that the polarization variations (e.g., \citealt{2015MNRAS.453.1669B}) during the flaring epochs can be observed.

As for the fact of that the fitting is not so good in high-energy bands as in low-energy bands, we propose an possible explanations for why the fitting is not so good in high-energy bands as in low-energy bands. It is believed that the extreme high-energy and/or high-energy synchrotron peaked BL Lac objects (EHBL or HBL) are considered as an origin of high-energy cosmic neutrinos (e.g., \citealt{2016NatPh..12..807K,2018ApJ...865..124M,2018Sci...361.1378I,2019MNRAS.483L..12C}). In this scenario, the acceleration and survival of ultrahigh-energy cosmic ray nuclei is possible in EHBL and HBL. As a consequence, the high energy photons and protons can induce cascaded processes resulting that the synchrotron and SSC emission components in the hard X-ray energy band (e.g., \citealt{2015MNRAS.447...36P}), and GeV $\gamma$-ray energy band (e.g., \citealt{2008ApJ...682..767Y}) are contaminated.

Both the solar observations and the numerical simulations (e.g., \citealt{2005ApJ...622.1251L,2010SoPh..266...71K}) have revealed that the magnetic reconnection is an inherently time-dependent, highly dynamic process. \cite{2009MNRAS.395L..29G} and \cite{2013MNRAS.431..355G} suggest the time-dependent aspects of reconnection may prove to be crucial in understanding blazar flares in terms of energetics and timescales. The present paper anticipates that the acceleration and loss processes experienced by the electrons in the blob may be strongly concentrated in the vicinity of the shock front. In this vicinity,  magnetic reconnection may result in exciting the time-varying EM fields, and strengthening the electrostatic acceleration (\citealt{2013ApJ...770..147C}). In this scenario, the assumed impulse variability of the EM fields can be expected.

\section*{Acknowledgements}
We thank the anonymous referee for valuable comments and suggestions.
This work was partially supported by the National Natural Science Foundation of China (Grant Nos. 11673060, 11763005, 11873043 and 11991051). Additional support was provided by the Specialized Research Fund for Shandong Provincial Key Laboratory (Grant No. KLWH201804), by the Key Laboratory of Particle Astrophysics of Yunnan Province (Grant No. 2015DG035), by the the Science and Technology Foundation of Guizhou Province (QKHJC[2019]1290) and by the Research Foundation for Scientific Elitists of the Department of Education of Guizhou Province (Grant No. QJHKYZ[2018]068).

\section*{Data availability}
The data underlying this article will be shared on reasonable request to the corresponding author.

\bibliographystyle{mnras}
\bibliography{article}

\label{lastpage}
\end{document}